# Linking Solar Minimum, Space Weather, and Night Sky Brightness

Albert D. Grauer*
Catalina Sky Survey
Lunar and Planetary Laboratory
University of Arizona, USA
adgrauer@email.arizona.edu

Patricia A. Grauer
Cosmic Campground International Dark Sky Sanctuary, New Mexico, USA

October 19, 2021

## Abstract

This paper presents time-series observations and analysis of broadband night sky airglow intensity 4 September 2018 through 30 April 2020. Data were obtained at 5 sites spanning more than 8,500 km during the historically deep minimum of Solar Cycle 24 into the beginning of Solar Cycle 25. New time-series observations indicate previously unrecognized significant sources of broadband night sky brightness variations, not involving corresponding changes in the Sun's 10.7 cm solar flux, occur during deep solar minimum.

New data show;
1) Even during a deep solar minimum the natural night sky is rarely, if ever, constant in brightness. Changes with time-scales of minutes, hours, days, and months are observed.
2) Semi-annual night sky brightness variations are coincident with changes in the orientation of Earth's magnetic field relative to the interplanetary magnetic field.
3) Solar wind plasma streams from solar coronal holes arriving at Earth's bow shock nose are coincident with major night sky brightness increase events.
4) Sites more than 8,500 km along the Earth's surface experience nights in common with either very bright or very faint night sky airglow emissions. The reason for this observational fact remains an open question.
5) It is plausible, terrestrial night airglow and geomagnetic indices have similar responses to the solar energy input into Earth's magnetosphere.

Our empirical results contribute to a quantitative basis for understanding and predicting broadband night sky brightness variations.  They are applicable in astronomical, planetary science, space weather,  light pollution, biological, and recreational studies.

# 1. Introduction

Natural night sky airglow is powered by solar energy received from space weather above and tropospheric activity below.[1] This complex environment is a unique laboratory providing a rich tapestry of solar and terrestrial phenomena. In the past, especially in the artificial light at night community there was little appreciation of the dynamic nature of the natural night sky. Awareness is growing that the study of natural terrestrial airglow may offer clues to researchers in fields as diverse as geomagnetic activity, climate change, high energy cosmic rays, planetary atmospheres, gravity waves, severe weather, light pollution, nocturnal plants and animals, recreation in parks, radio communications, and more. Scientific studies related to the brightness of the natural night sky are published in the astronomy, aeronomy, planetary, biological, meteorological, geomagnetic, and other forums of scientific and technical literature. Unfortunately, papers in these diverse fields do not reliably cross reference each other.

The intensity of the broadband night sky brightness determines the integration time to achieve the desired limiting magnitude and signal to noise for objects in a deep astronomical image.[2] A quantitative understanding of natural broadband night sky airglow is essential to discover faint Earth approaching asteroids and to schedule observations by telescopes like the Vera C. Rubin Observatory.[3,4] Astrophotography opportunities and stargazing world wide benefit from a knowledge of the broadband brightness of the natural night sky.[5] Mitigation of detrimental effects of artificial light at night (ALAN) require a knowledge of variations in the broadband brightness of the natural night sky.[6,7,8,9]

A broad range of atmospheric conditions impact astronomical observations from the Earth's surface.[4] In the past, many of the night sky brightness measurements, published in the astronomical literature, were made by research-grade instruments occasionally scheduled for this task or as a by product of another scientific research project.[10-15] Data and models of night sky brightness are used to plan observations and to protect observatory sites.[4,15,16] Astronomers have recognized that broadband night sky brightness has, diurnal, semi-annual, solar cycle variations.[9,15]

Researchers in the aeronomy community study physical and chemical processes in various layers of Earth's upper atmosphere which, in aggregate, account for the broadband airglow emissions we study in this paper. In 1970, Silverman published a comprehensive review of night airglow phenomena. It is still informative today.[17] Deutsch and Hernandez review the observational behavior of the intensity of the OI 558 nm line during geomagnetic quiet conditions.[18] Near Adelaide, Australia, 15 years of observations of atomic oxygen (OI) 558nm and hydroxyl (OH) 730 nm reveal annual, semiannual, solar cycle, and other variations in night airglow.[19] Researchers at Andes Lidar Observatory report 142 nights from September 2011 to April 2018 have an unusual patten of $O(^1S)$ night airglow enhancement with concurrent weaker OH(6,2) emission. This data set showed a semi-annual occurrence rate reaching maxima near the equinoxes.[20] Thirty years of airglow data at OI 557.7 nm and OI 630.0 nm have been used to establish a data driven model consistent with the GLOW airglow model. The result is the ability to visualize seasonal and solar cycle variations in red and green oxygen lines.[21,22] From 1998 to 2001, airglow emissions OI5577, O2b(0,1), and OH(6,2) and the rotational temperature of the OH band, at an equatorial location, show semiannual variations with maxima at the equinoxes and minima at the solstices.[23]

Geomagnetic scientists study the nature and frequency of geomagnetic storms, their relationship to events on the Sun, and their effects on the Earth's magnetosphere. Changes in natural night sky brightness , including auroras, provide clues about ionospheric and space weather processes. Extreme events in this realm have a significant impact on electrical grids, electronic communication, and navigation satellites. [24]

Geomagnetic activity has a long history of time series measurements and their analysis. Conversely, interruptions by the Sun, Moon, weather events, and the lack of a wide spread network of suitable measuring stations at natural night sky locations have left the study of broadband natural night sky airglow brightness variations relatively undeveloped.

The semi-annual variation in geomagnetic activity has been known for over 100 years.[25] In 1971, Russell and McPherron proposed a model to explain this phenomena in terms of the relationship between the z component of the interplanetary magnetic field, $[B(t)_z]_{GSM}$, and the z component of the Earth's magnetic field. [26] [both expressed in the Geocentric Solar Magnetospheric (GSM) coordinate system]. In the Russell and McPherron model the interaction between these two magnetic fields acts like a rectifier. When $[B(t)_z]_{GSM}$ is negative, opposite Earth's magnetic field, charged particles are more likely to penetrate the ionosphere. When $[B(t)_z]_{GSM}$ is positive, in the same direction as the Earth's magnetic field they are partially blocked. A negative $[B(t)_z]_{GSM}$ produces enhanced geomagnetic activity. Examples of the geomagnetic activity to which these authors refer include the number of geomagnetic storms per month and the geomagnetic index U. According to the Russell and McPherron model geomagnetic activity reached a maximum around 4 April 2019 and 7 October 2019 . Expressed in fractions of a year, F, these peaks are at F = 0.257 and F=0.769. It should be emphasized, this model is based on the statistics of geomagnetic events above a certain threshold. There are broad peaks and dips with a full width at half maximum of many days.

In a series of papers, Lockwood et al. [27,28,29] investigate the semiannual, annual, and Universal Time variations in the magnetosphere and geomagnetic activity. A key element in their research is an estimation of the power input into the magnetosphere, $P_\alpha$. They calculate $P_\alpha$, employing interplanetary measurements with the formula originally derived, theoretically, by Vasyliunas et al.. [30] $P_\alpha$ has only 1 free parameter, the coupling factor $\alpha$. It is driven by the speed, number density, ion mass measurements of the solar wind, and modulated by the interplanetary magnetic field's strength and orientation. In general, geomagnetic parameters have fractional variational amplitudes larger than corresponding temporal fractional changes in $P_\alpha$ . This amplification can be seen by comparing the am geomagnetic index with $P_\alpha$.[28] This research group shows the Russell-McPherron Effect is the principal driver of semi-annual geomagnetic activity even though it has a small impact on $P_\alpha$. Interestingly, they report the intensity of geomagnetic activity produced by the Russell-McPherron Effect is, apparently, amplified by the release of energy stored in the Earth's magnetospheric tail.

## 2. Data Collection and Analysis

In this paper, photometers are employed to obtain a time-series of differential photometric brightness measurements of the same place on the celestial sphere along the zenith declination relative to celestial sources. This procedure minimizes errors encountered using data from several different instruments and individual instrument drift in sensitivity if it exists.

Our research is enabled by accurate, low cost, scientific quality SQM-LU-DL [31] and TESS-W [32] photometers. They provide continuous measures of zenith night sky brightness dusk to dawn every night. These two photometers use the same detector, have slightly different fields of view, and different red responses.[31] The SQM-LU-DL uses a color filter. The TESS-W employs a dichroic filter. These filters and TSL237 photodiode they both employ sets the spectral response of each instrument. The SQM-LU-DL has a spectral response which spans most of the Johnson B and V filters while the TESS-W has a substantially greater red response spanning more than the Johnson-Cousins B,V, and R pass bands. [9] The SQM-LU-DL and TESS-W report instrumental $M(t)$ values in mag/arcsec$^2$. These two instruments are slightly different measures of broadband night sky brightness. However, their differential photometric measurements produce similar results. In the differential photometry mode we employ, our photometers are more accurate (error < 0.03 mag/arcsec$^2$) compared to when used as calibrated absolute photometers (error ~ 0.1 mag/arcsec$^2$). [31] The irradiance-to-frequency semiconductor detector employed by both the SQM-LU-DL and TESS-W instruments is calibrated to report the measurements in mag/arcsec$^2$. [33]

An approximate conversion of mag/arcsec$^2$ to cd/m$^2$ is:

$$L_v \text{ [in cd/m}^2] = L_0 * 10^{-(0.4 * M(t) \text{ [in mag/arcsec}^2])}$$

$L_0$ is $1.475 \times 10^5$ cd/m$^2$ in the AB System and $1.2216 \times 10^5$ cd/m$^2$ in the Vega system. [34]

On clear, astronomically dark nights, these single channel photometers, pointed at zenith, measure light accumulated from terrestrial airglow, stars, planets, scattered star light, zodiacal light, nebulae, galaxies, other faint astronomical sources, and anthropogenic skyglow, if present. SQM and TESS-W photometers sum all emissions in a broad cone to the edge of space over a relatively wide area of the sky. These characteristics make it difficult to identify the physical processes creating the emissions. These instruments produce time-series data to identify broadband airglow brightness events for further study.

Time-series data are collected at Cosmic Campground International Dark Sky Sanctuary (CCIDSS) and Catalina Sky Survey Mt. Lemmon Station (CSSMLS). Data from TESS-W photometers located at Spain Observatorio Astrofísico de Javalambre- Arcos de las Salinas/Teruel (Stars 18), Centre d'Observació del'Universe, Àger, Lleida, Spain (Stars 62), and Observatorio del Teide, Izaña, Tenerife, Spain. (Stars 211) were downloaded from the TESS Data Monthly data files using IAU-IDA format. [35]

| Site | Instrument | Latitude | Longitude | Elevation | Artificial Level [36] |
|---|---|---|---|---|---|
| CCIDSS | SQM-LU-DL | 33.4793° N | 108.9226° W | 1634 m | 0.632 µcd/m² |
| CSSMLS | SQM-LU-DL | 32.4420° N | 110.7893° W | 2791 m | 132 µcd/m² |
| Stars18 | TESS-W | 40.0371° N | 1.001815° W | 1589 m | 31.8 µcd/m² |
| Stars62 | TESS-W | 42.0246° N | 0.73479° W | 810 m | 37.0 µcd/m² |
| Stars211 | TESS-W | 28.2983° N | 16.5105° W | 2106 m | 123 µcd/m² |

Table 1: The sites and instrumentation used in this study.

The artificial levels in Table 1 are estimates from satellite data adjusted by ground-based SQM observations. [36] CCIDSS is a unique standard more than 60 km away from any significant source of artificial light. Analysis of all-sky images and the satellite estimate, of less than ½ % artificial light, indicates anthropogenic skyglow is unmeasurable at zenith at CCIDSS. [36,37]

We have developed techniques and software to process the time-series data. Our software selects individual instrumental time-series measurements, $M(t)$, taken at time t, when the Sun was more than 18 degrees below the horizon, the Moon was more than 10 degrees below the horizon, and the sky was clear. The 10° Moon limit is necessitated by large changes in lunar brightness as the Moon passes through its phases every month. On photometric nights, at the CCIDSS, measurements show that if the Moon is less than 74% illuminated it provides negligible light at zenith when it is 10° or more below the horizon. The situation changes rapidly when the Moon approaches the horizon. For example, a 19% illuminated Moon, 5° below the horizon, increases the zenith sky brightness by 0.02 mag arcsec$^2$. The Moon's zenith illumination, when it is still below the horizon, could be effected by moisture and/or dust in the atmosphere thus it is better to be conservative in selecting a value for this parameter. [31]

Sky clearness is measured by computing Chi Squared from a straight line fit to the data extending for 45 minutes on either side of the point in question. A Chi Squared of less than 0.009 for at least 1.5 hours rejects suspect data but not the rising Milky Way. [31] This metric is employed to exclude data from marginal non-photometric nights. As an additional check, the TESS-W near IR sensor is employed to estimate cloud cover. [32] At Stars 211, in only one case out of 198 nights, did the Chi Squared fit indicate clear skies, while the IR sensor indicated cloudiness. For the TESS-W data, the original 1 min data are averaged over 5 minute intervals to produce 5 minute samples. Our software calculates the Right Ascension (R.A.), Declination (Dec.) ,Julian Date(JD) , Local Sidereal Time (LST), Solar and Lunar Altitudes and other parameters for each $M(t)$ data point.

A total of 12,892, $M(t)$, time-series data points, sorted into ½ h bins in R.A., obtained at CCIDSS September 2018 through April 2020 are plotted in Fig. 1. The vertical distribution of sky brightness, at each sky position of R.A., is due to changes in terrestrial airglow. Since $M(t)$ are measured in mag/arcsec$^2$ brighter values are smaller numerically.

Plots similar to Fig. 1, for the other sites, observationally establish a quiescent value of airglow at each location on the celestial sphere. The existence of a minimum intensity of broadband airglow at each RA on the celestial sphere is a unique observational fact for each site. Alternatively, one could, also, establish a value for the quiescent airglow level by adding up the minimum values for all known sources of diffuse night sky brightness .[16]

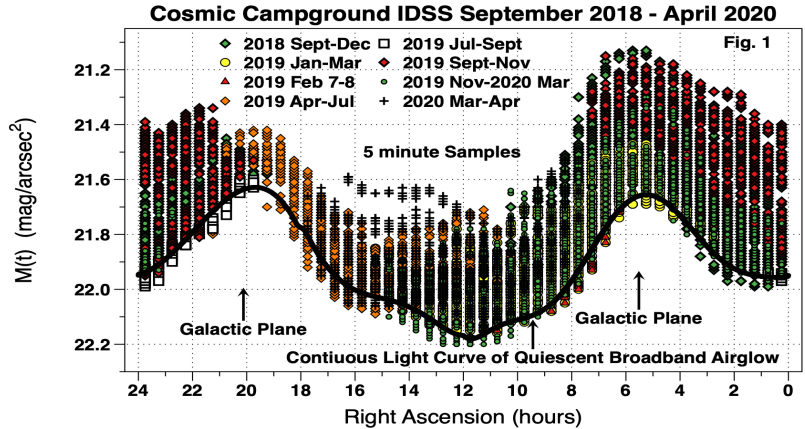

Fig. 1: The vertical axis is the instrumental sky brightness, M(t), n mag/arcsec$^2$. It shows changes in broadband night time terrestrial airglow at each sky position during various time periods as the data were accumulated. Since solar energy ultimately powers this airglow it is plausible observed broadband airglow variations are caused by changes in solar activity. The horizontal axis is the position in the sky in hours of R.A.( the data being sorted into ½ h bins). The minimum broadband night sky brightness at each location on the celestial sphere is an observational fact. The continuous light curve of quiescent airglow was calculated using the faintest 10% of the M(t) in each ½ h R.A. bin. The 10% is arbitrary and was chosen to have reasonable statistics. The fitted curve is indicated in Fig. 1. It is the broadband sky brightness when the terrestrial airglow is at minimum.

Along the zenith declination on the celestial sphere, ½ h bins in R.A. , are used as 48 standard candles. Each such standard candle is the average of the 10% faintest M(t) measurements at its location on the celestial sphere. A set of joined polynomials are fitted to the 48 standard candles to produce a continuous light curve of quiescent broadband airglow brightness. This fitted M(t) versus RA light curve is the background contribution from celestial sources and the broadband night airglow when it is at minimum. These concepts using data from CCIDSS are shown in Fig. 1. Each site has a unique light curve of quiescent broadband airglow brightness which depends on its latitude and the degree to which it is influenced by anthropogenic light. These light curves are an observational characteristic of each site. Each measured point's brightness above the quiescent airglow, ΔM(t), is obtained by subtracting the continuous light curve of quiescent broadband airglow from the data, point by point. This procedure removes light from the stars, planets, Milky Way, zodiacal light, other celestial sources, and constant anthropogenic light if present. Thus, each ΔM(t), is a differential photometric night sky brightness at time t, relative to the brightness of the same point on the celestial sphere when the airglow is at minimum. The same procedure is used to produce, ΔM(t), a time series of airglow brightness above its quiescent level for each site.

To evaluate conditions during the night at each site, we averaged the ΔM(t) data into ½ h time intervals relative to local midnight. The results are plotted in Fig. 2. The error bars, produced by real airglow variations, are +/- 1 standard deviation for the ½ h bin averages. At CCIDSS, the quiescent airglow light curve is relatively flat (average 0.18 mag/arcsec$^2$). The standard deviation of 0.136 mag/ arcsec$^2$ is produced by real changes in airglow. On long winter nights, before the onset of astronomical twilight, there does seem to be an increase of approximately 0.07 mag/arcsec$^2$ . The origin of this increase is unclear. There appear to be other reports of this phenomena in the literature.[38] It is plausible natural night sky broadband airglow, in certain situations, has a UT dependence similar to the geomagnetic indices. [29]

At CSSMLS, ΔM(t) is correlated with automobile driving patterns and scheduled outdoor lighting changes in and around Tucson, AZ. At Stars 18, Stars 62, and Stars 211 there appears to be prolonged morning and evening twilight when the Sun is more than 18° below the horizon. This result could be due to the extended red response of the TESS-W photometers and/or the European pattern of artificial lighting in nearby cities.

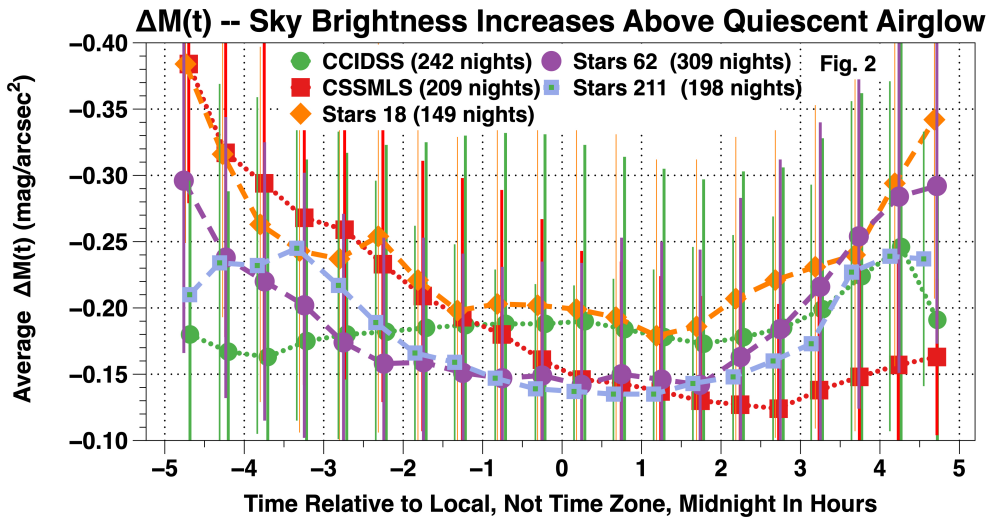

Fig. 2: The vertical axis is brightness of the ΔM(t) time series data averaged into ½ h time intervals relative to local midnight for each site. In the astronomical magnitude system of units the brightest ΔM(t) measurements are the most negative. The horizontal axis is the time in hours relative to local solar midnight.

For each site the ΔM(t) data versus the time relative to local midnight were fit to a quadratic function. Corrections using these functions were used to flatten the curves of Fig. 2, tighten the agreement between sites, and are small. In mag/arcsec$^2$, the site (median,stdev) correction values are CCIDSS (0.005,0.008), CSSMLS (-0.018,0.056), Stars 18 (0.017,0.048), Stars 62 (0.016,0.040), and Stars 211 (-0.033, 0.029). Each data point, ΔMC(t), is the corrected broadband airglow brightness above the continuous light curve of quiescent airglow brightness (Fig. 1). The resulting time-series of data points, ΔMC(t), for each site are used to track the natural zenith broadband night sky airglow uncontaminated by celestial and other sources above its quiescent level.

We compare the zenith natural broadband airglow above its quiescent level, ΔMC(t), during a deep solar minimum, with conditions in the near Earth environment. We employ solar wind data as compiled and presented on the NASA Omni Plus Browser [39], sunspot counts, 10.7cm(t) (2.8 GHz) solar flux, and Geomagnetic indices. [40,41,42,43]

We obtained the 1 h averages for $[B(t)_z]_{GSM}$, [Bz(t)], V(t) [Kp Speed], and n(t) [Kp-Proton Density] from the NASA Omni Plus Browser. [44] These parameters are the solar wind conditions at Earth's magnetic bow shock nose at time t.

Assuming protons are the predominate ions in the solar wind, we define a dimensionless measure of the solar wind kinetic energy, NKE(t), to be n(t) [measured Kp-Proton density] times v(t) [solar wind speed] squared divided by the median of this quantity for the period 4 September 2018 through 30 April 2020. Similar calculations were made to obtain dimensionless solar flux, N10.7cm(t), and dimensionless Potsdam Geomagnetic index, NAp(t). These dimensionless units are used to plot graphs which show relevant space weather conditions.

[B(t)z]GSM is the z component of the interplanetary magnetic field in geocentric solar magnetospheric system (GSM) in units of nT. In GSM system the z axis is aligned with the Earth's northern magnetic pole. Thus, a positive [B(t)z]GSM enhances the Earth's magnetic field while a negative [B(t)z]GSM opposes it.

An example of these concepts and calculations, Fig. 3, presents time-series data obtained on the night of 2019 February 7-8 (JD 2458522). Each data point, ΔMC(t), is the broadband airglow brightness above the continuous light curve of quiescent airglow brightness. This was a rare night when zenith sky brightness reached nearly record faint levels at all sites. In the direction away from the disk of the Milky Way (LST 11-14 h) the measured sky brightnesses were: CCIDSS (22.128 mag/arcsec², stdev = 0.016), CSSMLS (21.456 mag/arcsec², stdev = 0.009), and Stars 211 (21.469 mag/arcsec², stdev = 0.038). The photometers at CCIDSS and CSSMLS were in sync with one another for 4 h with a delta magnitude of 0.032 mag/arcsec² (stdev 0.014 mag/arcsec²). In the upper panel the dimensionless values of NKE(t), NAp(t), and N10.7cm(t) were all steady, at or below their median values. The [B(t)z]GSM had relatively small variations. All of these measurements taken together indicate a state of low solar and geomagnetic activity.

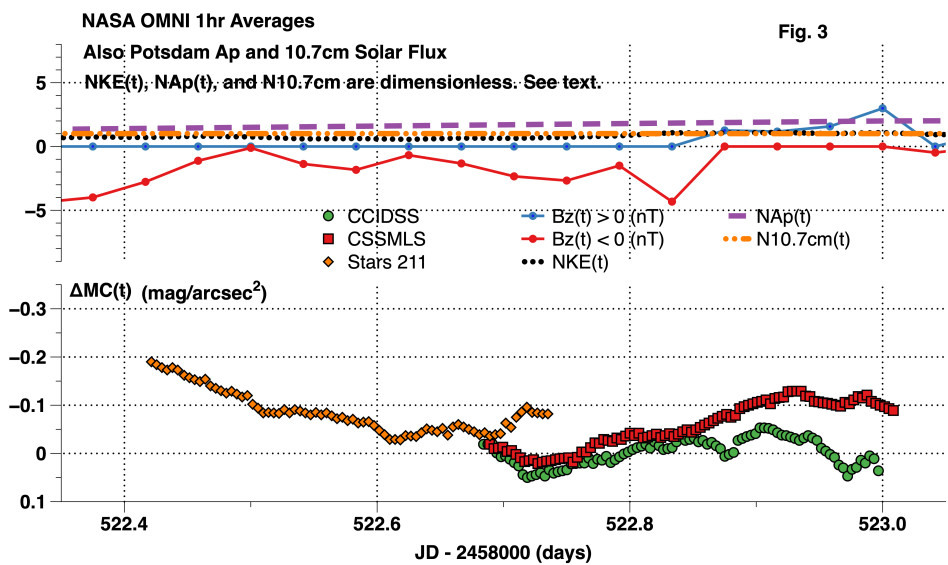

Fig. 3: A dark night experienced at sites spanning a distance of more than 8,500 km. The horizontal axis is the JD – 2458000 in days. [B(t)z]GSM in nT along with the dimensionless solar wind kinetic energy, NKE(t), dimensionless geomagnetic index, NAp(t), and the dimensionless radio solar flux, N10.7cm(t) are plotted in the upper panel to summarize space weather conditions. The time-series airglow brightness data points, ΔMC(t), in mag/arcsec² are plotted in the lower panel for Stars 211, CSSMLS, and CCIDSS. Each point is the differential photometric brightness relative to the continuous light curve of quiescent airglow brightness.

## 3. Results

Time-series photometric observations were taken at 5 sites spanning more than 8,500 km during the minimum of Solar Cycle 24 into the beginning of Solar Cycle 25. During this time one might have expected nightly airglow variations to be at a minimum. These data verify the natural night sky is rarely, if ever, constant in intensity. At a natural night sky location like CCIDSS during deep solar minimum the 5 min cadence time-series broadband airglow had a range in $\Delta MC(t)$ of 0.729 mag/arcsec$^2$. This corresponds to a maximum/minimum flux ratio of 1.957. The nightly airglow average $\Delta MCN(t)$ over a total of 241 nights at CCIDSS ranged from a very active night (2458792.979282JD) at -0.536 mag/arcsec$^2$ to a quiescent one (2458522.840410 JD) at 0.028 mag/arcsec$^2$. This nightly range of -0.564 mag/arcsec$^2$ corresponds to a flux ratio maximum/minimum intensity ratio of 1.68.

At CCIDSS, September 2018 through April 2020, 10 nights were recorded to have an average minimum broadband SQM brightnesses of 22.07 mag/arcsec$^2$ (stdev 0.03 mag/arcsec$^2$ ). These data were accumulated in the RA range 10.5 to 12.5 h.

Despite our instrument's dusk to dawn coverage every night, the observations were unavoidably interrupted by Sun, Moon, clouds, and instrument down time. For the total elapsed time during this research, 4 September 2018 through 30 April 2020, the % of time logged during clear astronomical dark conditions was for CCIDSS (7.4%), CSSMLS (5.6 %), Stars 18 (1.6 %), Stars 62 (3.3%), and Stars 211 (5.1%). Our data must be regarded as a small sample of sky brightness during this time. Thus, unless one has a worldwide network of monitoring stations, at dark sky locations, many important airglow events will be missed.

## 3a. Observations and 10.7 cm Solar Flux

The 948 nightly broadband airglow brightness measurements obtained from 37,437 individual time-series observations made at 5 locations, 4 September 2018 (2458365.5 JD) through 30 April 2020 (2458969.5 JD) are plotted in Fig. 4. The celestial and anthropogenic sources have been removed as outlined in Section 2 of this paper. The x axis is the time in Julian Date and the vertical axis is the nightly average of broadband airglow brightness, $\Delta MCN(t)$, above its quiescent level. In this paper $\Delta MCN(t)$ is defined to be the nightly average broadband airglow intensity above its quiescent level with the quadratic correction of Fig. 2. Since $\Delta MCN(t)$ are measured in mag/arcsec$^2$ brighter values are smaller numerically. The error bars are +/- 1 standard deviation. These standard deviations represent changes in broadband airglow during the night. Fig. 4 shows during a deep solar minimum natural night broadband airglow brightness varies by more than 0.5 mag/arcsec$^2$ (intensity ratio 1.58). Local night sky airglow brightness events span a distance of a few hundred km. Others can extend 8,500 or more km along the Earth's surface (see section 3e). Regular temporal gaps in the time series of $\Delta MCN(t)$ plotted in Fig. 4 are the result of the Moon having an elevation greater than 10 degrees below the horizon. These gaps, centered on full Moon, are approximately 29.35 days apart. The solar synodic rotation period is 26.24 days. The similarity between these two periods makes it difficult to identify the solar rotation in the data.

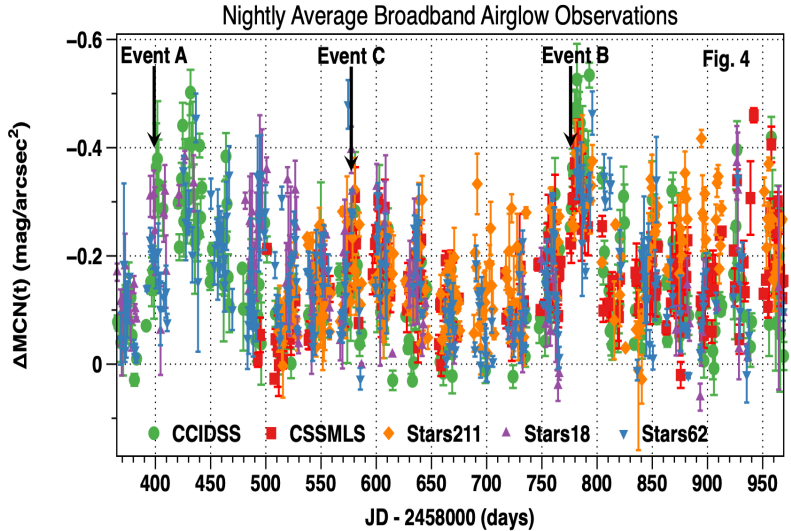

Fig. 4: The nightly average airglow above its quiescent level, $\Delta MCN(t)$, in ( mag/arcsec$^2$ ) for each site are plotted versus the time in Julian Date (days) . The error bars for each night are +/- 1 standard deviation about the mean. They represent changes during the night. Event A, Event B, and Event C. are discussed in the text. The actual measurement errors are less than 0.03 mag/arcsec.$^2$

Each nightly average broadband airglow measurement at CCIDSS, $\Delta MCN(t)$, was matched with the 10.7 cm flux value, from approximately 12 hr earlier (median 11.48 h). [45] This was done to compare the daytime solar illumination on the previous day with the night time airglow. The results are plotted in Fig. 5 . The broadband brightness changes we observed are not correlated with changes in solar EUV flux. During the entire period of our observations, the solar EUV measured by the 10.7cm(t) solar flux was at a low, relatively constant, level (average = 69.77 s.f.u., stdev = 2.45 , 1 s.f.u. = $10^4$ Jy). The range in this parameter during a solar cycle is from below 50 to above 300 s.f.u.. [46]

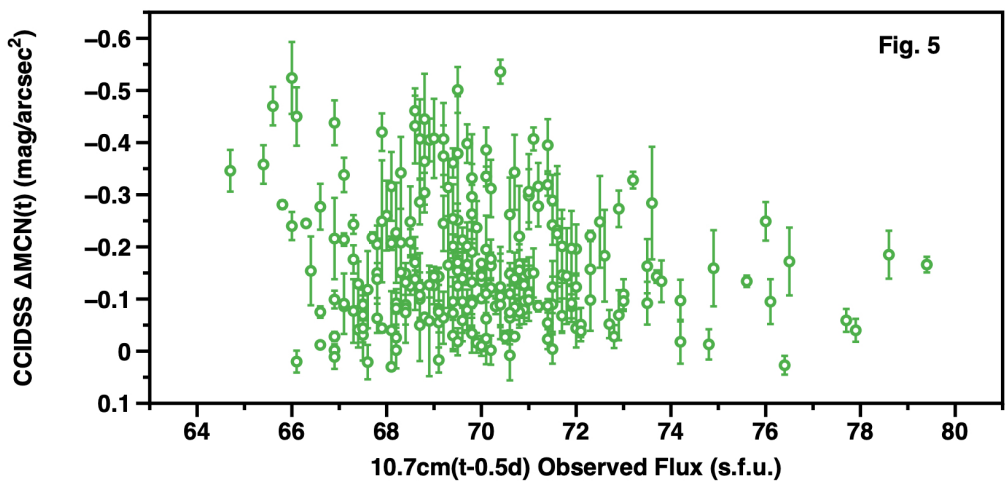

Fig. 5: The vertical axis is the average nightly broadband airglow at CCIDSS at time t, $\Delta MCN(t)$. The error bars are +/- 1 standard deviation about the mean. They represent changes during the night. The horizontal axis is the 10.7 cm flux value at time closest to  t - 0.5 d.

## 3b. Night Sky Brightness Increase Events

The ΔMCN(t), are corrected nightly average increases of the broadband night airglow above its quiescent level. These measurements are plotted versus Julian Date in Fig. 4. The error bars are +/- 1 standard deviation about the mean. They represent real changes during the night. Broad maxima in ΔMCN(t) occurred near JD 2458435 (12 November 2018), JD 2458589 (16 April 2019), and JD 2458786 (29 October 2019). These dates were obtained by fitting the ΔMCN(t) data to a quadratic formula extending 1 or 2 lunations either side of the peak. Each of these three broad maxima is an envelope of a number of broadband airglow brightness increase events separated by periods of bright moon. Events A, B, and C were selected as examples because they are representative and have corroborating data. They are marked in Fig. 4.

## 3b-1. Event A

From July to November 2018 the same large coronal hole was observed on the Sun.[47] It pointed in our direction every solar synodic period. After each such alignment, several days later Earth was engulfed in a high energy stream in the solar wind. These circumstances produced airglow and geomagnetic events world wide and was imaged by astronauts on the International Space Station. This series of events produced the broad maxima in night sky brightness near JD 2458435 (12 November 2018).

Event A is one of several broadband airglow increase events during an active period that spanned several solar rotations. Event A began at 2458398.895833 JD when a pulse in the dimensionless solar wind proton kinetic energy, NKE(t), was observed at Earth's bow shock nose. At this time our planet encountered an energetic stream in the solar wind. Before Event A the space weather conditions were relatively steady. The upper panel of Fig. 6 plots some space weather parameters before and after Event A. The dimensionless solar wind kinetic energy, NKE(t), increased to more than 4 times its median level. NAp(t), the daily dimensionless Potsdam geomagnetic Ap index increased to more than 5 times its median value. The peak Ap on day 24598399 JD was 56. It increased to more than 5 times its median level, indicating a significant geomagnetic disturbance. Meanwhile, the solar EUV as measured by the 10.7cm(t) radio flux remained near its median value of 70 SFU.

In the lower panel of Fig. 6 the x axis is the time in Julian Date – 2458000 (days) and the vertical axis is the nightly average of broadband airglow brightness, ΔMCN(t), at CCIDSS, above its quiescent level (mag/arcsec²). The error bars are +/- 1 standard deviations about the mean. They are real changes during the night. The time-series data for individual nights are plotted in Fig. 8.

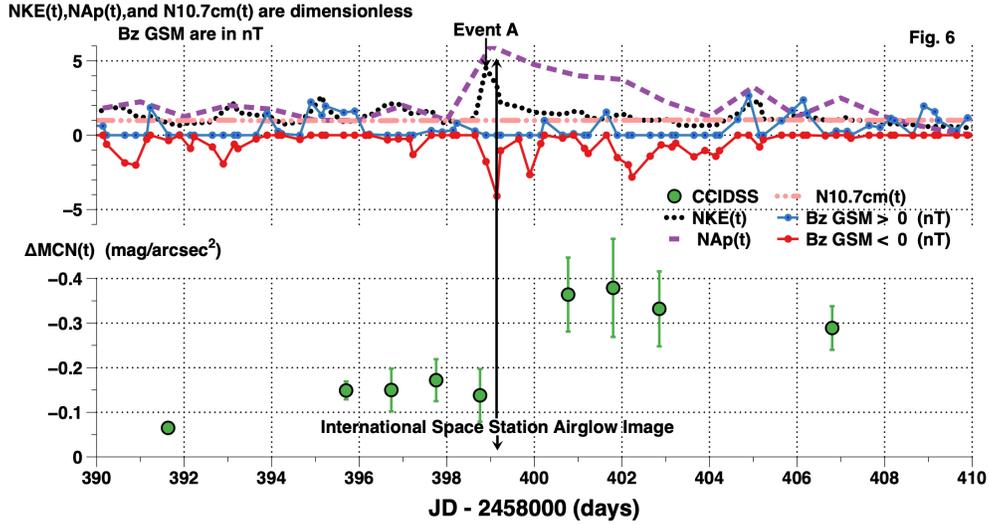

Fig. 6 (upper panel): Space weather conditions NKE(t), Nap(t), and N10.7cm(t) are the dimensionless kinetic energy, Potsdam geomagnetic index Ap(t), and 10.7cm(t) solar radio flux. See the text Section 2 for definitions of these dimensionless quantities. [B(t)z]GSM measurements in nT are plotted in red when Bz GSM < 0 and blue when Bz GSM > 0.
Fig. 6 (lower panel): Broadband average nightly airglow measurements,, ΔMCN(t), at CCIDSS above their quiescent levels are plotted versus Julian date. The +/- 1 standard deviation error bars indicate significant changes during the night. See Fig. 8 for each night's data plotted individually.

At 2458399.15646 JD an astronaut on the International Space Station took an image showing Earth engulfed in a bright orange airglow.

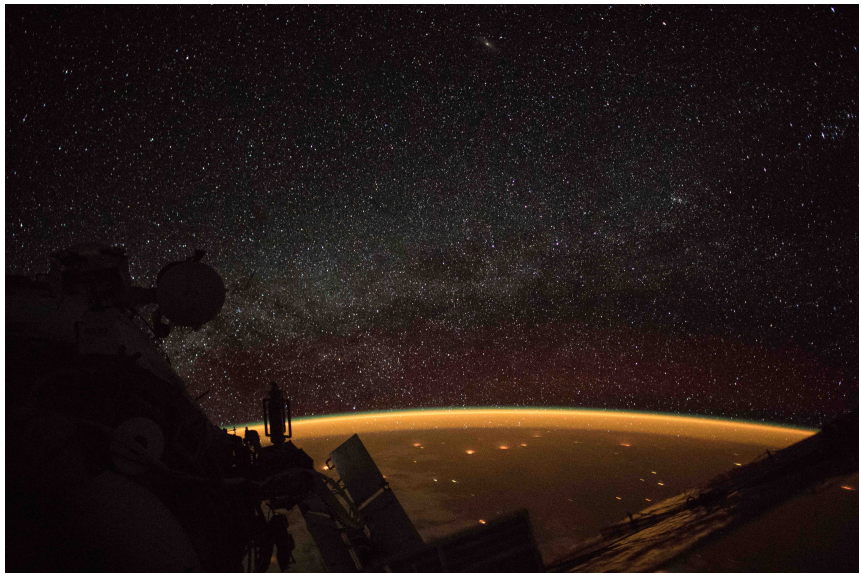

Fig.7: NASA International Space Station astronauts took the Image ISS-057-35382 on 2018.10.07 at 15:45:17.99 (JD =2458399.15646) location 30.6S, 133.4E. NASA issued a press release stating the Earth was surrounded by airglow.

The development of broadband night sky brightness Event A at CCIDSS is shown in Fig. 8. At Earth's bow shock nose, Event A, began with the arrival of a pulse in the dimensionless solar wind proton kinetic energy, NKE(t), at 2458398.895833 JD. The broadband 5 min cadence, time-series airglow measurements , ΔMC(t), above their quiescent levels are plotted versus the time in hours relative to local midnight for 7 nights in Fig. 8. These nights are designated by XXX the Julian Date being 2458XXX JD. (XXX = 396,397,398,400,401,402, or 406). Before Event A, the nights 396, 397, and 398 showed relatively small variations in broadband airglow brightness during the night with maxima near local midnight. This is typical for broadband airglow during a relatively quiet night. The night of 397 peaked at approximately 0.5 h before local midnight at maximum brightness, ΔMC(t), of -0.23 mag/arcsec$^2$ . The nights of 396 and 397 tracked each other for 5.34 h with average difference of 0.022 mag/arcsec$^2$ (stdev = 0.020 mag/arcsec$^2$). 396 and 397 were remarkably stable broadband airglow nights. The gap in the data from the night of 398 was caused by a band of clouds which passed overhead. After Event A the night sky brightness continued to increase reaching a maximum on night 401. The nights of 402 and 406 showed decreasing broadband airglow brightness levels as space weather conditions returned to more normal levels.

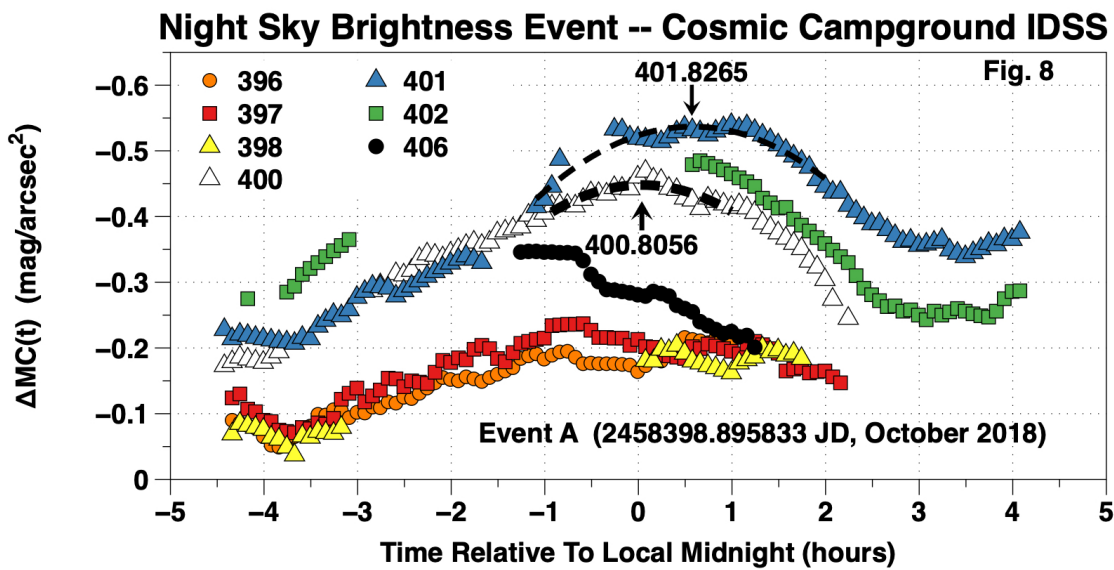

Fig. 8: The 5min cadence, broadband time-series airglow observations, ΔMC(t), at CCIDSS are plotted versus the time in hours relative to local midnight. Each ΔMC(t) point is the differential photometric broadband airglow brightness relative to the continuous light curve of quiescent airglow brightness (Fig. 1). This plot shows the shock wave Event A produced a broadband airglow intensity disturbance which lasted for several days. The triggering Event A began near 2458398.895833 JD.

## 3b-2. Event B

During the summer and fall of 2019 large coronal holes repeatedly pointed toward Earth. This series of events produced a broad maxima in night sky brightness near JD 2458786 (29 October 2019) in Fig.4 . Large increases in broadband night sky brightness occurred when particularly energetic streams in the solar wind impacted Earth's magnetosphere. Event B is one of several broadband airglow increase events during an active period that spanned several solar rotations. It is marked on Fig. 4.

Zoltán Kolláth's image (Fig. 9) was captured at CCIDSS on 20.10.2019 at 03:53 UT (2458776.66181 JD) at the beginning of an extended night sky brightness episode.[37] Fig. 10 shows the image's temporal relationship to the other data. In Fig. 9, Green (558nm) oxygen and orange (589nm) sodium airglow are visible over the entire sky. The R,G, and B channels in the digital camera data provide estimates of the strength and spatial structure of the oxygen and sodium lines. [37]

Mackovjak et al. used the AMON UV photometer and all sky camera to observe the airglow 557.8 nm (O1-green line), 568.5 nm (no airglow), 630.0 nm (OI-red-line), and 700-900 nm (OH) from 20 October 2019 to 1 November 2019. A plot of these data from JD 2458777 to 2458783 has been published.[48] The Mackovjak et al. data show a series of airglow increase brightness events in the time before and after Event B.

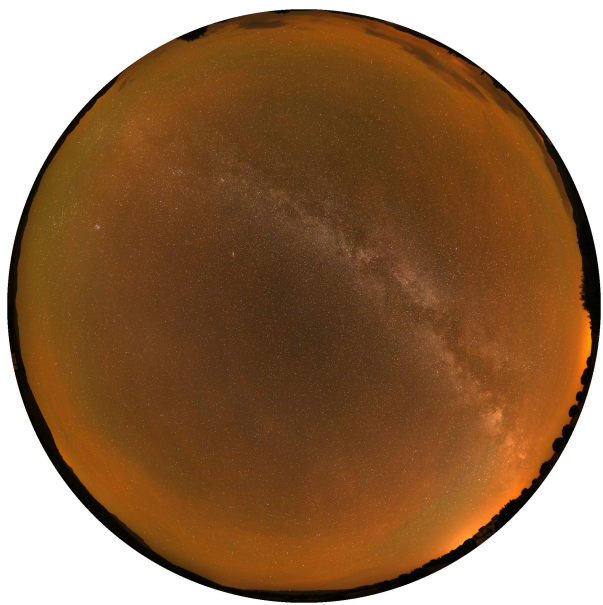

Fig. 9: On the night of October 19-20, 2019 Zoltán Kolláth captured this true color image at CCIDSS as part of a survey of dark sky sites in the American southwest. North is at the top of the image. He reported the airglow was unusually bright that night. His simple estimate gives 1.3*dsu* increase in the red channel by the sodium line and 0.40-0.45*dsu* by the sodium and the 558 nm oxygen line in the green channel of his DSLR. [37]
 These data are in agreement with the SQM-LU-DL data simultaneously recorded at CCIDSS

Space weather parameters and night airglow time-series brightness observations from JD 2458774 (19 October 2019) to JD 2458789 (1 November 2019) are plotted in Fig. 10. The upper panel is a plot of normalized solar wind and geomagnetic parameters versus the time in JD (days). The lower panel is a plot of the 5 min cadence broadband time-series airglow observations above the quiescent level, $\Delta MC(t)$, for each site in mag/arcsec$^2$ versus the time in JD - 2458000 (days). There was a slow rise and fall in the airglow brightness over a two week interval. This period of time was characterized by predominately negative $[B(t)_z]_{GSM}$ and a NKE(t) which varied significantly above its median value. Event B began near 2458780.8333 JD (24 October 2019) when a high energy stream in the solar wind produced a shock wave at Earth's bow shock nose. A pulse in NKE(t), nearly 5 times its median level, deposited energy into the Earth's magnetosphere. This produced a dramatic increase in geomagnetic activity, NAp(t). Meanwhile, the solar EUV as indicated by the N10.7cm(t) flux was low and constant and is uncorrelated with broadband airglow brightness changes. The period of negative $[B(t)_z]_{GSM}$ which followed Event B allowed energetic charged particles to penetrate deep into Earth's ionosphere. The shock wave triggered large variations in airglow brightness during the night reaching peak brightnesses near local midnight (Fig. 11). Of interest is the broadband airglow increase before the shock wave arrived. Perhaps, the shock wave triggered a release of stored magnetospheric energy in the near-Earth tail as has been demonstrated for geomagnetic events.[29] This plot defies a simple explanation.

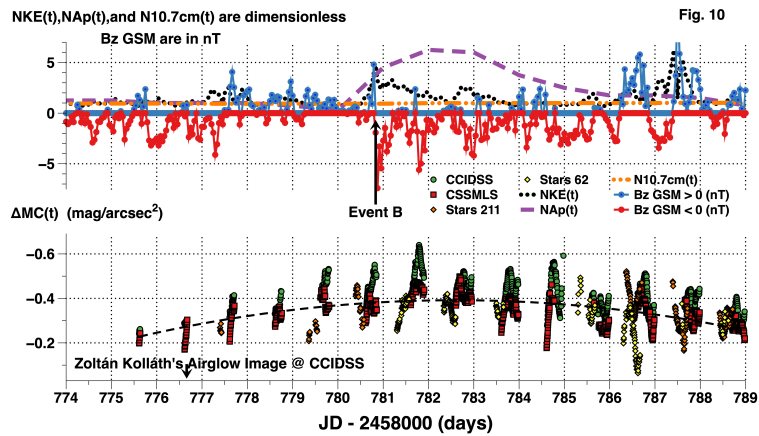

Fig. 10 (upper panel): The normalized geomagnetic index, NAp(t), normalized solar wind kinetic energy, NKE(t), normalized N10.7cm(t) solar radio flux along with the positive and negative z components of the interplanetary magnetic field in the GSM coordinate system in nT are plotted versus Julian Date (JD) in days. The shockwave in the solar wind arriving at 2458780.8333 JD features a dramatic sign change in $[B(t)z]_{GSM}$ and pulse in NKE(t) which appears to trigger geomagnetic and airglow variations. Similar events are observed on other occasions.
Fig. 10 (lower panel): The 5 min cadence broadband night time-series airglow measurements above the quiescent level, $\Delta MC(t)$, for each site are plotted versus Julian Date (JD) in days. Of interest is the broadband airglow increase in brightness before the arrival of the shockwave. This suggests preconditioning of the magnetosphere-ionosphere system may allow a shock wave to trigger a broadband night sky brightness increase event. Data from CCIDSS and CSSMLS, physically separated by 209 km, tracked together 6 days before and 8 days after Event B. During the intervening period they deviated significantly from each other. This may indicate broadband airglow features smaller than 209 km.

page-15

The nightly evolution of Event B, initiated by a shock wave at 2458780.8333 JD, is shown in Fig. 11. The broadband 5 min cadence, time-series airglow measurements, ΔMC(t), above their quiescent levels are plotted versus the time in hours relative to local midnight for 6 nights in Fig. 11. Before Event B the space weather conditions were not steady (see Fig. 10). The Event B shock wave triggered a geomagnetic disturbance whose peak daily Ap(t) was more than 6 times it's median value. The Ap(t) peak values for the nights 781,782, and 783 were 32, 56, and 39 indicates a significant geomagnetic event occurred.

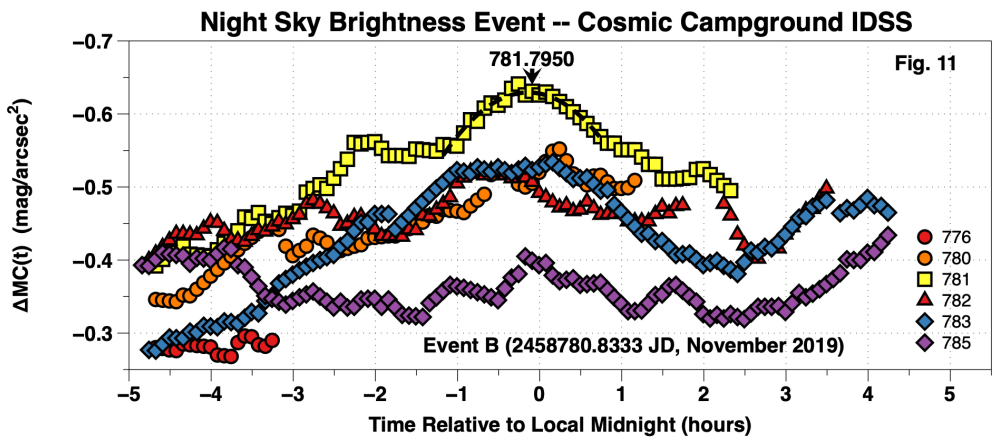

Fig. 11: The 5 min cadence broadband time-series airglow, ΔMC(t), at CCIDSS above its quiescent level are plotted versus the time in hours relative to local midnight. The ΔMC(t), measurements are plotted for the nights of 2458XXX where XXX is 776,780,781,782,783, and 785. The triggering Event B began near 2458780.8333 JD.

From JD 2458777 to 2458783 the solar EUV as measured by the 10.7cm(t) flux was low and constant (average 65.9 s.f.u., stdev =1.9).

### 3b-3.  Event C

March-April 2019 nightly airglow averages, ΔMCN(t), are compared with satellite, geomagnetic, and solar flux observations in Fig. 12 .   This time range was first identified by the effect night broadband airglow increases had on a large-scale municipal street light dimming experiment. Data show the midnight sky brightness over Tucson was correlated to changes in the natural airglow observed at CCIDSS. [6]  The plot, Fig. 12,  reveals a widespread increase in night sky airglow during a time when  $[B(t)_z]_{GSM}$ was predominately negative. Negative values of the z component of the interplanetary magnetic field diminish Earth's magnetic field allowing charged particles from the Sun to penetrate the Earth's magnetic shield.  The normalized solar wind kinetic energy, NKE(t) varied significantly above its median level and NAp(t) show enhanced geomagnetic activity. These quantities are defined in Section 2 of this paper.   Enhanced broadband airglow coincides with Russell-McPherron prediction of increased geomagnetic activity.[26]   This effect can, also, be seen in Fig. 6 and Fig. 10.

page-16

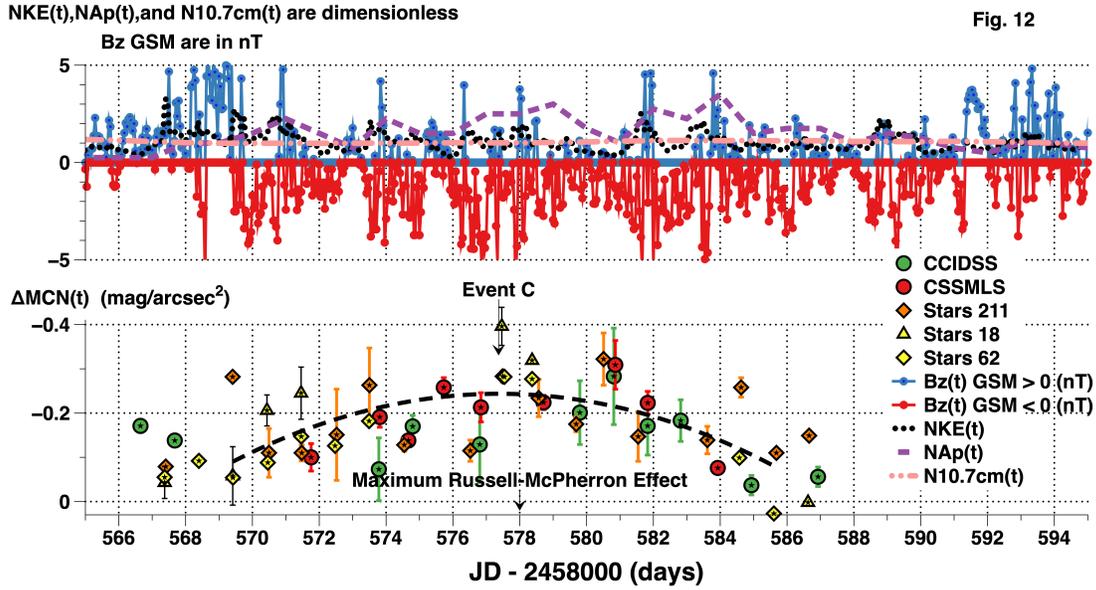

Fig. 12: The upper panel plots measures of solar and geomagnetic activity parameters versus time in JD. The lower panel plots nightly airglow brightness averages above the quiescent level, ΔMCN(t), versus JD. The error bars are +/- 1 standard deviation. They represent real brightness variations during the night. The actual measurement errors are less than 0.03 mag/arcsec$^2$. The dashed line is a quadratic fit to all the data. The negative values of [B(t)z]GSM (plotted in red) allowed charged particles from the solar wind to penetrate Earth's magnetic shield enhancing geomagnetic activity and terrestrial airglow. The Russell-McPherron effect models a statistical enhancement of geomagnetic activity over a long time base. Additional broadband night airglow data are required to firmly establish night airglow is statistically predictable using the Russell-McPherron effect.

## 3c. Semi-Annual Brightness Variations

The distribution of 948 nightly airglow averages, ΔMCN(t), obtained at five sites, 4 September 2018 through 30 April 2020 are plotted in Fig. 13. On the vertical axis, each point is the corrected nightly average of airglow brightness, ΔMCN(t) above its quiescent level. The celestial and anthropogenic sources have been removed as outlined in Section 2 of this paper. The horizontal axis is the time in fractions of a year (F). The data were sorted into 36 bins with an F width of 0.0278. The smoothed data curve was obtained from the 36 data bins using a 5 point triangular weighting function. The smoothed, binned, data curve of Fig. 13 shows a semi-annual variation in airglow brightness with broad peaks near 0.273 F and 0.837 F. The peak near 0.837 F has an amplitude and location strongly influenced by high speed streams in the solar wind from coronal holes on the face of the Sun which impact the Earth's magnetosphere (sections 3a and 3b). The amplitude of the unperturbed broadband airglow semiannual variation during solar minimum is ~ 0.2 mag/arcsec².

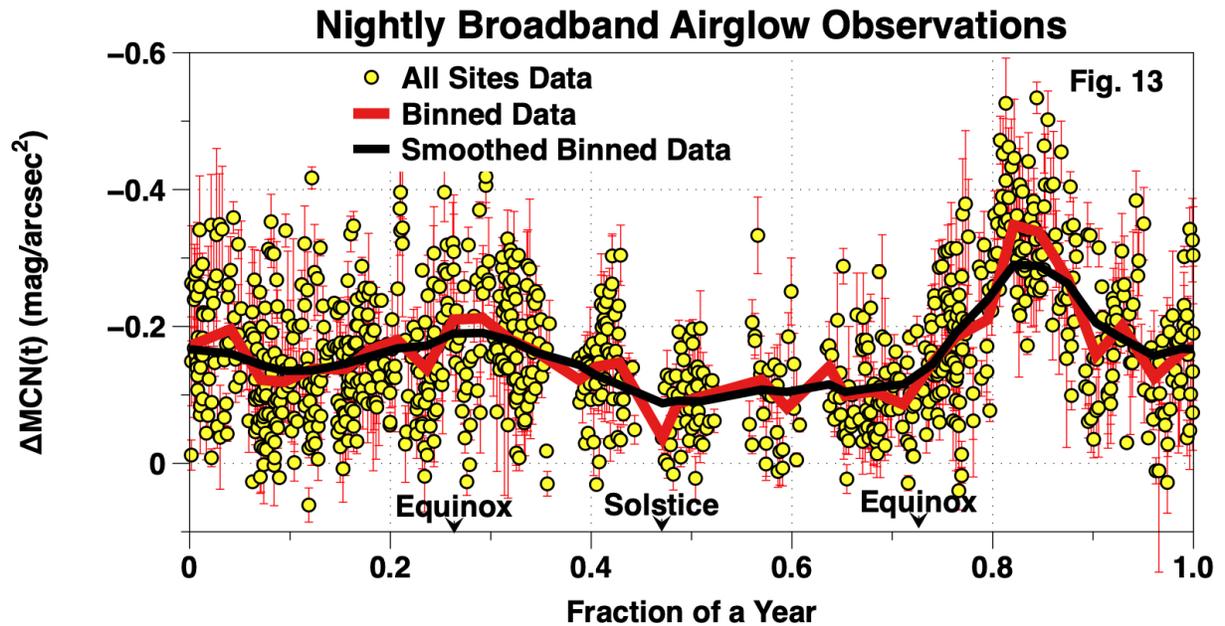

Fig. 13: The vertical axis are the ΔMCN(t) nightly average airglow data from Fig. 4. All 5 sites are plotted with the same symbol. The binned data and smoothed binned, data are described in the text. The horizontal axis is F, the fraction of a year. The error bars are +/- 1 standard deviation and represent real changes during the night. This plot clearly shows the semi-annual variation in night sky brightness. Given the random nature of solar events, the precise amplitude and the location of the maximums and minimums are likely to shift as the temporal time base increases in length. This wave form is similar to the am geomagnetic index.[28]

### 3d. Bz GSM

A total of 14,800, 1h cadence, $[B(t)_z]_{GSM}$ data points for the period JD 2458365.00 to JD 2458990.458 (3 September 2018 through 20 May 2020) were obtained from the NASA OMNI website.[44] This is a very noisy data set. It has a maximum of 14.3 nT and a minimum of -14.5 nT. The median is -0.17 nT with a standard deviation of 1.98. Binned $[B(t)_z]_{GSM}$ data points are plotted in Fig. 14 versus the fraction of year F. To accomplish the binning, , the 1 hr cadence OMNI $[B(t)_z]_{GSM}$ data were divided into 72 bins each 5.069 days wide in fractions of a year (F). The $[B(t)_z]_{GSM} > 0$ are the blue points and the $[B(t)_z]_{GSM} < 0$ are the red points for each bin. The error bars are +/- 1 standard deviation and show the distribution of $[B(t)_z]_{GSM}$ binned data. The Events A, B, and C all occurred when the average $[B(t)_z]_{GSM}$ was less than zero. This plot shows that near the Vernal Equinox $[B(t)_z]_{GSM}$ is negative for a sustained period of time. Near the Autumnal Equinox strong Events A and B influenced $[B(t)_z]_{GSM}$ during that period of time. It should be emphasized that the NASA OMNI $[B(t)_z]_{GSM}$ measurements at Earth's bow shock nose have a chaotic nature since they are dominated by energetic streams and shock waves in the solar wind.

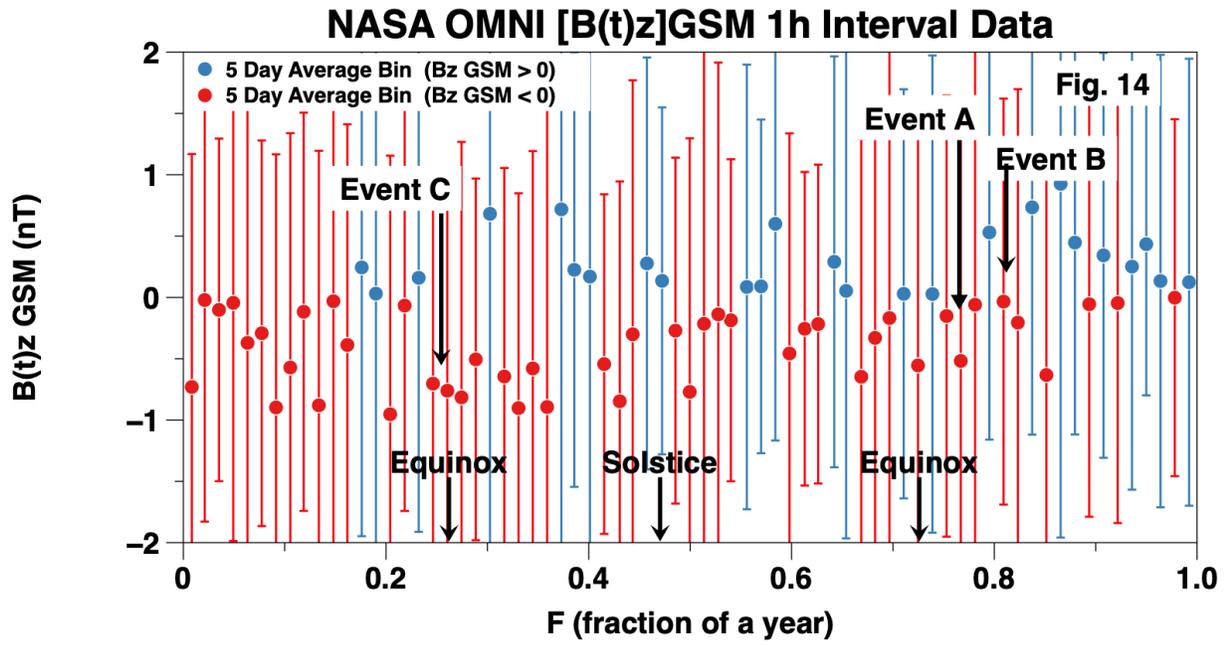

Fig: 14: The NASA OMNI 1 hr $[B(t)z]$GSM data. Plotted on the vertical axis are 5 day averages with standard deviations. The horizontal axis is fractions of a year. A sustained period during which $[B(t)z]$GSM < 0 occurred near the vernal equinox.

## 3e. Geographic Variations

Nights with unusually intense airglow detected through broadband filters as well as others with unusually faint airglow detected through broadband filters were observed at locations further than 8,500 Km apart. To illustrate these phenomena we calculated an average night sky brightness ΔMCN(t) with standard deviation for each of the 948 nights observed at CCIDSS, CSSMLS, Stars 18, Stars 62, and Stars 211.

CCIDSS and CSSMLS had 123 clear nights in common while CCIDSS and Stars 211 had 107 clear nights in common. These data were sorted to produce (x,y) pairs for each night in common. The x,y pairs, (CCIDSS,CSSMLS) and (CCIDSS, Stars 211) for each night in common are plotted as individual points in Fig. 15. The results for Stars 18 and Stars 62 are similar but not included since they are at a more northern latitude.

The (CCIDSS,CSSMLS) points in Fig. 15 track together more tightly than do the (CCIDSS, Stars 211) points. On a number of nights the broadband airglow at Stars 211 was significantly brighter than it was on the same night at CCIDSS. However, one of the very brightest night sky airglow nights at CCIDSS correspond to some of the very brightest airglow nights at both CSSMLS and Stars 211.

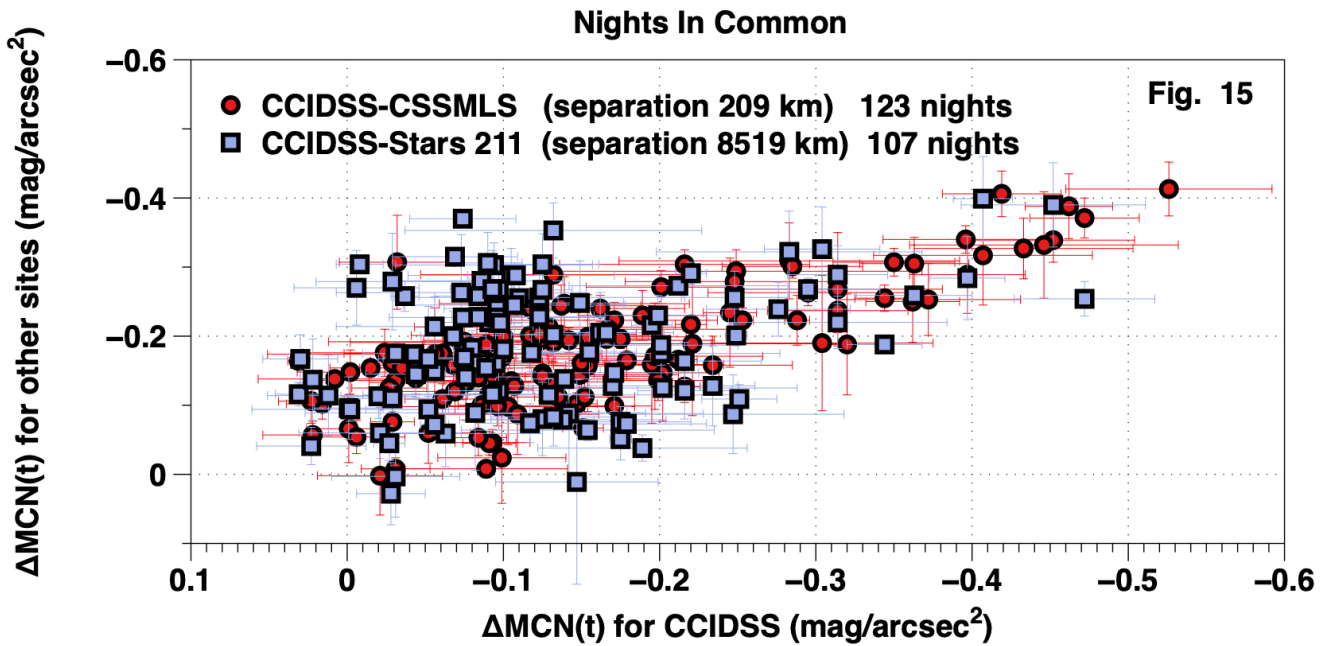

Fig. 15: Clear nights in common for CCIDSS-CSSMLS and CCIDSS-Stars 211. The horizontal axis is the ΔMCN(t) [nightly average] and standard deviation for CCIDSS. The vertical axis is the ΔMC-N(t) [nightly average] and standard deviation for CSSMLS (round red points) and the ΔMC-N(t) and standard deviation for Stars 211 (square blue points).

## 4. Discussion

There is no simple cause and effect relationship between solar activity, space weather, and changes in broadband night terrestrial airglow. However, there are coincidental relations between events on the sun, conditions in the near Earth environment, and the brightness of the night sky which need to be explored.

This paper is based on detailed analysis of individual astronomically dark nights during a deep solar minimum. Taking a completely different approach, Alarcon et al. used statistical methods to analyze data from dozens of TESS-W photometers during solar minimum.[9] Many of their instruments are in locations with more than 0.5 mag/arcsec$^2$ of human caused artificial illumination. They report short time-scale variations in the night sky airglow which they attribute to events in the mesosphere.

Variations in broadband night sky airglow are not always accompanied by changes in 10.7 cm solar flux (Fig. 5). To put this fact into prospective, relationships between the brightness of the natural night sky and solar activity as measured by the 10.7 cm radio flux are discussed by many authors. Krisciunas et al. obtained V-band sky brightness from Cerro Tololo Inter-American Observatory CCD images over the course of a solar cycle.[12] A comparison of Fig. 5, this paper, and Fig. 2 from the Krisciunas et al. paper indicates during solar minimum there are substantial variations in broadband airglow brightness when the 10.7 cm solar flux is relatively constant. In the same paper Krisciunas et al. Fig. 5 present a linear relationship between CCD V-band sky brightness observations and the average of the 10.7 cm solar flux measured 4.5 days earlier over a complete solar cycle.[12] The range of these CCD broadband night sky brightness data correspond to approximately a (nL, s.f.u) of from (55 nL, 70 s.f.u.) to (80 nL, 215 s.f.u). These data indicate a V-band increase of -0.41 mag/arcsec$^2$ corresponds to an increase of in 10.7 cm solar flux of 145 s.f.u over the course of a solar cycle. Observations of the oxygen green line, 5577 nm, were made at Sacramento Peak, New Mexico from January 1960 to June 1963. The ($I_{557.7nm}$ in Raleighs), 10.7cm in s.f.u) were found to range from approximately (60 Raleighs, 78 s.f.u.) to (440 Raleighs, 170 s.f.u.).[49] These data show a change of 92 s.f.u corresponds to an increase of approximately 2.2 magnitudes in the brightness of the green line. These and other studies show physical processes for which night airglow intensity is strongly coupled to the solar EUV as measured by the 10.7 cm solar flux. Rosenberg and Zimmerman are able to correlate intensity of the 557.7 nm (OI) line and the 10.7 cm Solar flux changes during a solar cycle.[49] A recurrent hypothesis is, changes in photoionization on the Earth's day-side produce subsequent variable night time airglow by re-radiation from atoms and molecules in a complex chemical environment. This concept can be used to relate changes in solar EUV, estimated from the 10.7cm solar flux, to daily and yearly night sky brightness changes throughout a solar cycle. Our Fig. 5 indicates on some occasions other physical processes produce significant changes in the intensity of broadband night airglow.

The existence of a semiannual variation in the brightness of the broadband natural night airglow has been reported in the astronomical literature. Patat reviews this situation and presents new data derived from spectroscopic observations in the UBVRI passbands at the Cassegrain focus of the 8.2 m telescopes at the Paranal Observatory in Chile.[15] These observations were made during the decline from maximum to minimum of Solar Cycle 23. Patat reports a clear seasonal variation in the broadband VRI passbands with two broad maxima (April-May and October) and two broad minima (July-August and December-January). In the aeronomy literature, observations obtained over many years show there are semiannual oscillations in OI 558 nm and OH 730 nm.[18,19] TIMED/SABER observations show a global distribution of oscillations in OH night glow emission with semiannual, annual, and quasi-biennial time scales.[50] Our data show during a deep solar minimum there are semiannual broadband night airglow variations with maxima near the equinoxes (Fig. 13). Our observations cover only a 1.65 year long period. As a result the location of the maxima and minima in Fig. 13 are shifted by space weather events such as Event A and Event B. It should be emphasized; in every case the semiannual modulation in night sky airglow is derived from noisy data. The precise details of the semiannual variation likely require a long base line in time.

Is it plausible the semiannual variations in night sky air glow are driven by similar processes to those explained by the Russell and McPherron model for the statistical behavior of geomagnetic events?[26] Quantities derived from NASA OMNI $[B(t)_z]_{GSM}$ data are plotted versus fraction of a year (F) in Fig. 14 . It should be emphasized NASA OMNI data are measurements which include both the effects of space weather and the changing magnetic field orientations beautifully illustrated by Lockwood et al. in their figures 3 and 4.[27] Our Fig. 14 shows $[B(t)_z]_{GSM}$ to be predominately negative for a substantial period of time near the vernal equinox while we were accumulating data. The situation is less clear near the autumnal equinox. In the Russell-McPherron effect the chaotic, highly variable, solar wind is modulated by the interplanetary magnetic field to produce a statistically discernible geomagnetic activity pattern using data encompassing a number of years. The data presented in Fig. 13 and Fig. 14 suggest it is possible broadband night sky airglow variations follow from changing magnetic field alignments. Data from more years will be required to put this hypothesis on a solid statistical basis.

For broadband night sky airglow increases observed near Event A and Event B:
1) There are substantial changes in broadband night airglow brightness.
   (Fig. 4, Fig. 6, and Fig. 10).
2) The Earth encounters energetic streams in the solar wind. (Fig. 6 and Fig. 10)
3) The 10.7cm solar flux is low and constant near its minimum values (Fig. 6 and Fig. 10).
4) $[B(t)_z]_{GSM}$ is predominately negative (Fig. 6 and Fig. 10).
5) Images show a strong orange tint (Fig. 7 and Fig. 9).
6) The time development takes place over several nights. The peak brightness occurred
   several days after the triggering event. (Fig. 8 and Fig. 11).
7) There are substantial variations in broadband airglow during the night
   (Fig. 8, Fig. 10, and Fig.11).
8) The large variations in broadband airglow during the night at various sites are
   apparently correlated (Fig. 10).
9) There are large amplitude geomagnetic events (Fig. 6 and Fig. 10).

Event A, Fig. 8 shows a airglow brightness change of approximately 0.40 mag/arcsec$^2$ (1.44 times in intensity units) . Fig. 10 for Event B the airglow brightness also changes by more than 0.35 mag/arcsec$^2$ (1.38 times in intensity units). Alarcon et al. used statistical methods to analyze data from dozens of TESS-W photometers.[9] They report observing short time-scale variations on most nights which they attribute to airglow events in the mesosphere. They do not report on the details of any specific events.

The orange tint of the images presented in Fig. 7 and Fig. 9 suggest Na airglow may have been excited during Event A and Event B.[37]

For broadband night sky increase Event C:
1) There are relatively small broadband night airglow brightness increases over a wide geographic area (Fig. 12).
2) $[B(t)_z]_{GSM}$ is predominately negative for a significant period of time (Fig. 12 and Fig. 14).
3) The 10.7cm solar flux is low and constant near its minimum value (Fig. 12).
4) Solar and geomagnetic activity varies above their median values (Fig. 12).

Some airglow night sky events are global. An example is Fig. 7. Reed et al. obtained maps of night airglow at 630 nm from Ogo 4 satellite observations about a year before the time of maximum of Solar cycle 20. They show some very large structures in night sky airglow at mid-latitudes which change during the night.[51] Alarcon et al. used statistical methods to analyze data from dozens of TESS-W photometers.[9] They find a correlation between the physical separation of photometers and the standard deviation of the differences in their measurements. The data plotted in Fig. 15 show some very bright broadband airglow nights at CCIDSS are also very bright on the same night at both CSSMLS and Stars 211. Similarly some of the nights at CCIDSS during which the night sky broadband airglow was faintest were also, on the same night, among darkest nights at both CSSMLS and Stars 211. The observations plotted in Fig. 15 suggest the presence of broadband night sky airglow events of various dimensions or, perhaps, they demonstrate semiannual variations for stations at similar latitudes. The amplitude of semiannual broadband night sky airglow variations shown in Fig. 13 as well as those published in the literature are on the order of 0.2 mag/arcsec$^2$.[9,15] The range of variations observed in Fig. 15 is much larger than these measured semiannual amplitudes of broadband night sky airglow. There are a number of nights which have bright broadband night airglow at Stars 211 and are simultaneously dim at CCIDSS. These nights do not seem to support the semiannual hypotheses. Even so, the reason for the trend displayed in Fig. 15 remains an open question.

Data presented in this paper show variations in broadband airglow with time-scales on the order of minutes, hours, days, and months. There are, undoubtedly, a number of interacting physical processes which cause what is observed. Localized aurorae outside the auroral oval have been observed and have an origin which is not completely understood.[52] We show during solar minimum the variations in broadband airglow are unrelated to changes in the 10.7 cm solar radio flux. Instead, they are likely caused by charged particles from the solar wind which enter Earth's magnetosphere. Observations by NASA's IMAGE spacecraft and the joint NASA/European Space Agency Cluster satellites show that huge cracks develop in the Earth's magnetosphere for hours allowing charged particles from the solar wind to enter the ionosphere.[51] Some of these cracks appear on a seasonal basis and others present themselves in a more random fashion. They are associated with magnetic reconnection of Earth's magnetic field lines with those in the interplanetary magnetic field. It is postulated that magnetic fields from the Sun and Earth reconnect on Earth's day side. From there the solar wind transports the reconnected magnetic flux to Earth's night side where it is stored in the magnetospheric tail. This stored energy can be released by a triggering event.[53,54,55] The data from Events A and B, qualitatively, match this scenario (see Figs. 6,8,10, and 11). Event C implies broadband night time airglow around the vernal equinox follows the Russell-McPherron prediction for geomagnetic activity (please see Figs. 12,13, and 14).

Broadband night sky airglow and geomagnetic activity are both likely responses to changes in space weather.  Sounding of the atmosphere by the SABER instrument aboard the NASA TIMED satellite over the course of a solar cycle relate changes in thermospheric cooling with variations in solar ultraviolet irradiance and geomagnetic activity. [56]  Solar ultraviolet irradiance and geomagnetic processes are, also, related to cooling of the thermosphere by infrared radiation from nitric oxide over the duration of a solar cycle.[57]  Thirteen years of data from the SABER instrument find the intensities of four night glow emissions are strongly coupled to solar radiation.[58]  In this paper, Events A and B show broadband night sky airglow increase events coincident with Earth interacting with an energetic stream in the solar wind. Event C documents observed night sky broadband airglow brightness variations coincident with the alignment of the z component of Earth's magnetic filed and the interplanetary magnetic field. Our results contribute to the continuing effort to unravel how solar-terrestrial interactions modulate night sky airglow emissions.

## 5. Conclusions

New observational data reveal changes in night time airglow which are significant in broadband astronomical and artificial light at night studies. SQM and TESS-W photometers sum all emissions in a broad cone to the edge of space and over a wide area of the sky.  These characteristics make it difficult to identify the physical processes creating the emissions. However, these instruments produce time-series data which can be used to identify broadband airglow brightness events for further study.

During deep solar minimum the broadband night sky airglow is never constant in intensity. For our data set, there were nights when the SQM broadband airglow intensity at the natural night sky location CCIDSS, became as faint as 22.07 mag/arcsec$^2$ .  On other nights the SQM broadband airglow was brighter than 21.57  mag/arcsec$^2$.

We report, during solar minimum, significant episodes of increased night sky airglow are not produced by changes in 10.7 cm solar flux.  We find these night sky brightening events are coincident with:
   1) Changing orientation of the interplanetary magnetic field relative to Earth's magnetic field and
   2) Earth entering streams of energetic solar wind.
It is plausible episodes of increased broadband night sky airglow we observe could be amplified by a release of energy stored in Earth's magnetospheric tail triggered by a shock wave in the solar wind.

Sites more than 8,500 km along the Earth's surface experience nights in common with either very bright or very faint night sky airglow emissions. The reason for this observational fact remains an open question.

Our data suggests the terrestrial night airglow responds to the energy input into the Earth's magnetosphere in a fashion similar to the geomagnetic indices.

We strongly advocate the establishment of a global network of photometers located in places where anthropogenic skyglow is at a minimum. These instruments would be used to track brightness variations of the natural night sky.   Established astronomical observatories are the places to start.  These measurements will have a significant impact on the studies of astronomy, space weather,  light pollution, biology, and recreation.

## 7. Acknowledgements

Eric Christensen, John Barentine, Richard Green, Anthony Tekatch, Greg Leonard, Hannes Gröller, and Simon Mackovjak provided information, encouragement, editing suggestions, and advice. Zoltán Kolláth captured and analyzed a true color image, on a night of enhanced airglow, at CCIDSS. The program DataGraph for macOS by Visual Data Tools, Inc. was used to produce the figures and make the curve fitting calculations in this paper.

Jaime Zamorano and Rafael González Fuentetaja provide essential support to the TESS Data Monthly data files using IAU-IDA format.

Dr. Natalia E. Papitashvili maintains and provides assistance for the NASA Omni Website. It is a valuable source of information concerning the Solar-Terrestrial interaction at the Earth's bow shock region.

## 8. Author Contributions (names must be given as initials)

PAG and ADG began this research to certify the Cosmic Campground International Dark Sky Sanctuary in southwestern New Mexico.

ADG is responsible for manuscript content and data analysis.

PAG is responsible for manuscript content and editing changes to make the document coherent.

## 9. Additional Information (including Competing Interests Statement)

The authors declare no competing interests.

**Materials and Correspondence.**
Corresponding Author
Albert D. Grauer
PO Box 2143
Silver City, NM 88062
575-590-7536
algrauer@mac.com

**Data Availability Statement**
The data that support the findings of this study are available on request from the corresponding author [ADG].